\documentstyle[12pt]{article}
\newcommand{\vev}[1]{\langle #1 \rangle}
\begin{document}
\noindent
\hspace*{10.5cm}Feb. 1995\\
\hspace*{10.5cm}OS-GE 44-95\\
\hspace*{10.5cm}hep-ph/9502381

\vspace*{2cm}
\begin{center}
  \begin{Large}
    On the Phase of $B^0$--$\bar {B^0}$ Mixing Matrix Elements\\
    in the SUSY Standard Model\\
  \end{Large}

\vspace*{1.5cm}
{\large T.~Kurimoto}\footnote{e-mail: krmt@phys.wani.osaka-u.ac.jp\\
\hspace*{.6cm}address after April 1995: Department of Physics, Faculty
of Science,\\
\hspace*{4.5cm}
Toyama University, Gofuku 3190, Toyama 930, Japan}\\
\vspace*{.7cm}
Department of Physics, Faculty of Science,\\
Osaka University,\\
Machikaneyama 1-16, Toyonaka, Osaka 560, Japan\\
\vspace*{3cm}
{\bf Abstract}
\end{center}

It is shown that the complex phase of $B^0$--$\bar {B^0}$
mixing matrix element is same as that of the standard model
one up to the minor correction of the order of $(m_c/m_t)^2$ or
less in the SUSY standard model.
 This conclusion is
valid as far as the Yukawa coupling constants are perturbative and
realistic, and can be available for other realistic models where
generation mixing and CP violation are solely due to
Yukawa interaction among quarks and Higgs doublets.

\newpage

Investigation of CP violation phenomena in $B$ physics is
very important not only for fixing the parameters in the
standard model but also for exploring new physics beyond
the standard model.\cite{bpr} CP violation in $B$ decay
can occur through the interference between  $B^0$--$\bar {B^0}$
mixing and a decay amplitude as
in the case of decay rate asymmetry in neutral $B$ meson decay
into a CP eigenstate like $J/\psi K_s$. There, the complex phase of the
$B^0$--$\bar {B^0}$ mixing matrix element,
$M_{12} = \langle B^0|{\cal H}^{\Delta B=2}|\bar {B^0}\rangle$,
is significant since the asymmetry is proportional to
$ {\rm Im}[(q/p)\rho]$, where
\begin{equation}
  \frac{q}{p}\cong \frac{M_{12}^*}{|M_{12}|}, \qquad
\rho \equiv \frac{Amp[\bar {B^0} \rightarrow f_{CP}]}{
                  Amp[ B^0 \rightarrow f_{CP}]},
\end{equation}
in the $B$ meson case.
The standard model gives\cite{SMB}
\begin{equation}
  \arg [M_{12}] \cong \arg [(V_{td}^* V_{tb})^2],
\end{equation}
where $V_{ij}$ is the Kobayashi-Maskawa matrix element.\cite{KM} If
a new physics contributes significantly to $B^0$--$\bar {B^0}$ mixing,
the mixing matrix element $M_{12}$ is different from
the standard model value.
The discrepancy between the standard model prediction and the
experimental data can shed light on new physics search.
The magnitude of $M_{12}$ is related to the oscillation
frequency of  $B^0$--$\bar {B^0}$ mixing or the mass difference
between two mass eigenstates, $\Delta M_B$, while the phase
to CP violation asymmetry.
The mass difference has been already
measured, and the CP asymmetry will be within our reach by
the $B$ dedicated experimental facilities now under
construction.\cite{bfac}
Investigation of the prediction on $B$ physics in models
beyond the standard model is now called for the search
of new physics at those $B$ experiments.

Here it is shown that the phase of $M_{12}$ is same as
the standard model value in the models which satisfy
these conditions;
\begin{enumerate}
\item CP violation and generation mixing are solely originate
from Yukawa couplings among quarks and Higgs $SU(2)_L$ doublets.
\item Yukawa coupling constants are perturbative and realistic.
\item Quarks are given in 3 generations.
\item Flavor changing neutral current is forbidden at the
tree level. (Natural flavor conservation\cite{NFC})
\end{enumerate}
As a concrete example we first investigate SUSY standard model,
which satisfies above conditions. Then we extend the results
to other models like multi-Higgs models where the above conditions
are satisfied.
The CP asymmetry is same as the standard prediction but the magnitudes
of $B^0$--$\bar {B^0}$ mixing can differ in these models. Then the
so called unitarity triangle\cite{bpr}  made from angles alone
and that from sides alone are not consistent with each other, which
is a sign of new physics.

It has been shown that the SUSY standard model gives
the $B^0$--$\bar {B^0}$ mixing matrix elements
$M_{12}$ of the same phase as that of the standard model one in
the case where top quark Yukawa coupling is far larger than
other Yukawa couplings.\cite{SUSYB}. Recently this is found to
be the case  even if top quark
Yukawa coupling is not so larger than bottom quark Yukawa coupling
as far as the Yukawa couplings are perturbative and
realistic by Goto, Nihei and Okada's numerical computation.\cite{OKG}
Here we give an analytic explanation of it.

SUSY standard model has 2 Higgs doublets, $H$ and $H'$, which couple to
quark fields as
\begin{equation}
  {\cal L}_{\rm Yukawa} =
  \overline{D_R}y_D Q_L H + \overline{U_R}y_U Q_L H'
                          +{\rm h.c.}, \label{eqn:ly}
\end{equation}
where $Q_L = (U,D)_L$, and we abbreviated 3 generation quark fields
by $U=(u,c,t)$ and $D=(d,s,b)$. The Yukawa coupling constants, $y_D$
and $y_U$, are $3 \times 3$ complex matrices.
The key point of the following argument lies in the fact that
generation changing interaction is solely controlled by the above
Yukawa interaction ({\it condition} 1). It is also important to
pay attention to the fact that left-handed quark fields always
come right-side of the Yukawa couplings while right-handed fields
to left-side.
Without loss of generality
we can take the basis where $y_D$ is real positive diagonal and $y_U$
is  expressed in terms of up-type quark masses and Kobayashi-Maskawa
matrix  $V$ ({\it condition} 4);
\begin{eqnarray}
  y_D &=& diag(m_d,m_s,m_b)/\vev{H^0} \equiv {\tilde M}_D\\
  y_U &=& diag(m_u,m_c,m_t)V /\vev{H'^0}\equiv {\tilde M}_U V
\end{eqnarray}
The coupling among quarks and $W$ boson is still
generation diagonal in this basis. Then any generation mixing in
the theory should be expressed in terms of the Yukawa couplings
({\it condition} 2).
For example, the d-type scalar quark mass matrix,
\begin{equation}
 (  \begin{array}{c c}
   \tilde {D_L}^{\dag} & \tilde {D_R}^{\dag}
  \end{array} )
  \Biggl[
    \begin{array}{cc}
         (L-L)_D & (L-R)_D \\
         {(L-R)_D}^{\dag} & (R-R)_D
    \end{array}
  \Biggr]
  \Biggl(
  \begin{array}{c}
   \tilde {D_L} \\
    \tilde {D_R}
  \end{array}
  \Biggr),
\end{equation}
is expanded in the Yukawa couplings in the following way;
\begin{eqnarray}
  (L-L)_D &=& m_{DL}^2 I +
          \sum_{{\bf p}=(p_1,p_2,\dots ) } \kappa _{DL}^{({\bf p} )}\
          ({y_D}^{\dag}y_D)^{p_1}({y_U}^{\dag}y_U)^{p_2}\cdots
         ({y_D}^{\dag}y_D)^{p_{j}} \nonumber \\
         &=&  m_{DL}^2 I +
           \sum_{{\bf p}=(p_1,p_2,\dots) } \kappa _{DL}^{({\bf p} )} \
           {{\tilde M}_D}^{2p_1}(V^{\dag}{\tilde M}_U^{2p_2} V)\cdots
             {{\tilde M}_D}^{2p_j},\nonumber\\
  (L-R)_D &=& m_{DLR}^2 I +
        \sum_{{\bf p}=(p_1,p_2,\dots) } \kappa _{DLR}^{({\bf p} )} \
           {{\tilde M}_D}^{2p_1}(V^{\dag}{\tilde M}_U^{2p_2} V)\cdots
             {{\tilde M}_D}^{2p_j} {\tilde M}_D,\nonumber \\
(R-R)_D &=& m_{DR}^2 I +
        \sum_{{\bf p}=(p_1,p_2,\dots) } \kappa _{DR}^{({\bf p} )} \
           {\tilde M}_D
           {{\tilde M}_D}^{2p_1}(V^{\dag}{\tilde M}_U^{2p_2} V)\cdots
             {{\tilde M}_D}^{2p_j} {\tilde M}_D.\nonumber \\
       & &
\end{eqnarray}
The index $p_i$ takes a value of non-negative integer. The
coefficients $\kappa^{({\bf p} )}$ and the generation diagonal
parts are constants fixed
in the model. Note the order of Yukawa couplings given in these
formulae as pointed out below Eq.(\ref{eqn:ly}).

We calculate the contribution to $B^0$--$\bar {B^0}$ mixing
by treating the generation mixing as a perturbation (mass
insertion). The gauge interactions are all generation
diagonal in our basis. The generation mixing part of
the gluino contribution to $b\rightarrow d$ transition shown in
Fig.1 is given as
\begin{equation}
  [{{\tilde M}_D}^{r_1}(V^{\dag}{\tilde M}_U^{2r_2} V)
  {{\tilde M}_D}^{r_3}\cdots (V^{\dag}{\tilde M}_U^{2r_{n-1}} V)
    {{\tilde M}_D}^{r_n}]_{13}.
\end{equation}
By using the hierarchy among quark masses
\begin{equation}
  m_t \gg m_c \gg m_u, \quad m_b \gg m_s \gg m_d
\end{equation}
and that among
Kobayashi-Maskawa matrix elements ({\it conditions} 2 and 3),
\begin{equation}
  V = O \left(
  \begin{array}{ccc}
     1 & \lambda & \lambda^3 \\
      \lambda & 1 & \lambda^2 \\
     \lambda^3 & \lambda^2 & 1
  \end{array}
   \right), \quad (\lambda = |V_{us}|)
\end{equation}
we find that the dominant contribution is
given by
\begin{eqnarray}
  & & ({{\tilde M}_D}^{r_1})_{11}(V^{\dag})_{13}
      ({\tilde M}_U^{2r_2})_{33} (V)_{33}
  \cdots (V^{\dag})_{33}({\tilde M}_U^{2r_{n-1}})_{33} (V)_{33}
    ({{\tilde M}_D}^{r_n})_{33} \nonumber \\
  &\propto & m_d^{r_1}m_t^p m_b^q |V_{33}|^{2s}
(V^{\dag})_{13}V_{33}
  =  m_d^{r_1}m_t^p m_b^q  |V_{tb}|^{2s}V_{td}^*V_{tb} .
\end{eqnarray}
and the next leading part is suppressed by $(m_c/m_t)^2$.
This situation is shown schematically in Fig.2. Therefore we get
\begin{equation}
  \arg [M_{12}^{gluino}]\cong \arg [(V_{td}^* V_{tb})^2]
                        \cong \arg [M_{12}^{SM}].
\end{equation}
Similar calculation of chargino, neutralino and charged Higgs
contributions give the same conclusion although the calculation
is a bit complicated.

Now we give a simpler explanation of the above result.
We have at most 4 types of $b \rightarrow d$ transition
effective operators under the conditions 1 and 2 ;
\begin{equation}
  \begin{array}{ll}
     \overline{d_L} \Gamma [y^{\dag}\cdots y]_{13} b_L, &
    \overline{d_R} \Gamma [y_D (y^{\dag}\cdots y)]_{13} b_L, \nonumber \\
     \overline{d_L} \Gamma [(y^{\dag}\cdots y)y_D^{\dag}]_{13} b_R ,&
    \overline{d_R} \Gamma [y_D (y^{\dag}\cdots y)y_D^{\dag}]_{13} b_R ,
  \end{array}
\end{equation}
where $\Gamma$ is a combination of $\gamma$ matrices, and
$y=y_D$ or $y_U$. So the leading coefficient of the above terms is
given by
\begin{equation}
   m_d^r m_t^p m_b^q  |V_{tb}|^{2s}V_{td}^*V_{tb}
\end{equation}
in our basis of Yukawa coupling
under the conditions 3 and 4.
We find that
\begin{equation}
  M_{12} = \langle B^0|f (\bar d \Gamma b)
                        (\bar d \Gamma 'b)
           |\bar {B^0}\rangle
         \cong (V_{td}^*V_{tb})^2 \times {\rm (real \ factor)}.
\end{equation}
This conclusion holds as far as the model satisfies the four
conditions given before because we have
made no model specific calculation.

Before closing this letter we give one comment on the power
of quark masses in $M_{12}$. The standard model also satisfies
the four conditions. If we calculate $M_{12}$ treating the
generation changing by mass insertion method as given before
in the SUSY standard model case, the contribution shown in
Fig.3 gives $M_{12}\propto m_t^4 (V_{td}^*V_{tb})^2$. But we
know that $M_{12}\propto m_t^2 (V_{td}^*V_{tb})^2$ in the ordinary
calculation if $m_t \ll M_W$.\cite{INL} This mystery is solved
once we include all the remaining contributions of mass insertion.
The exact formula of $M_{12}$ in the standard model depends on
quarks masses as follows;
\begin{eqnarray}
 & &   \xi_x \xi_y [
     -\frac{3}{4}xy + \frac{x^2 \ln x - y^2 \ln y}{x-y}
        +O(m_q^6)] \label{eqn:num}\\
 & \rightarrow &
     \xi_x^2 [-\frac{3}{4}x^2 +x + 2 x \ln x  +O(m_q^6)]\qquad
 (y \rightarrow x),
\end{eqnarray}
where $\xi_i = V_{id}^*V_{ib}$ and $x=(m_x/M_W)^2$.
The numerator of the second term of Eq.(\ref{eqn:num}) is $O(m_q^4)$.
Inclusion of full mass-insertion contribution gives
the denominator $x-y$, and the ln term appears if careful treatment of
infra-red divergence is done in the calculation.
In any case the phase of $M_{12}$ is controlled
by $(V_{td}^*V_{tb})^2$ because
$|m_t^n(V_{td}^*V_{tb})^2 |\gg |m_c^n(V_{cd}^*V_{cb})^2|$ as
far as $n \ge 1$.

In conclusion we have shown that the phase of the $B^0$--$\bar {B^0}$
mixing matrix element is unique in the models where generation mixing
and CP violation  solely originate from the naturally flavor conserving
Yukawa coupling among 3 generations of quarks and Higgs doublets
as far as the Yukawa coupling constants are realistic and perturbative.
Examples of the models which satisfy these criterions
are the standard model, multi-Higgs models with NFC and
the SUSY standard model.
We need informations both of sides and angles of the unitarity
triangle to find a signal of these models at coming $B$ factories.

\vspace*{1cm}
\noindent
{\it Acknowledgment}

The author would like to thank Professor Y.~Okada at KEK
for fruitful discussions. This work is supported in part
by Grant-in Aids for Scientific Research from the Ministry
of Education, Science and Culture (No. 06740218 and 05228104).

\newpage

\noindent
\begin{Large}
  {\bf Figure Captions}
\end{Large}

\noindent
\begin{description}
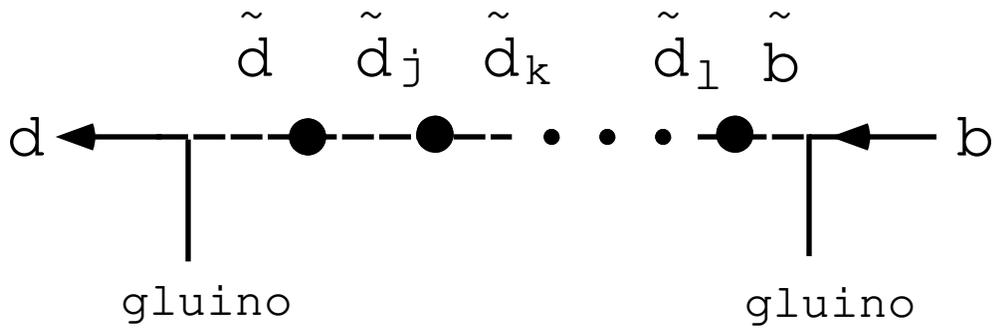
\item[Fig.1] Gluino contribution to $b\rightarrow d$ transition.
The black circle represents a scalar quark mass matrix element.
The indices  $j$, $k$ and $l$ specify scalar quarks.
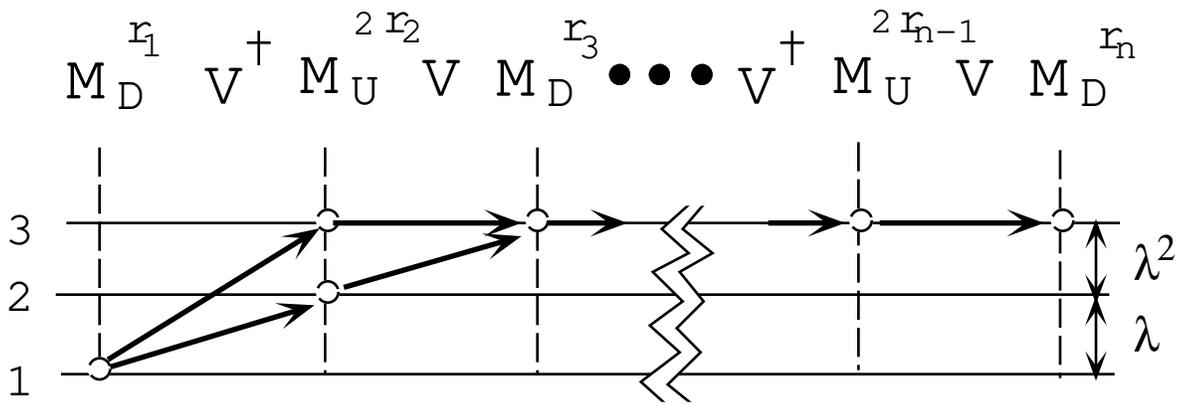
\item[Fig.2] Diagrammatic estimation of the matrix element (8).
The number at the left side represents generation. Transition
between  1st and 2nd (2nd and 3rd) generations gets $O(\lambda)$
($O(\lambda^2)$) factor from quark mixing matrix $V$. A factor
of a power of quark mass is multiplied at each circle. At least
the factor of $O(\lambda^3)$ is necessary in $b\rightarrow d$ transition.
The dominant contribution is obtained by multiplying as least
1st or 2nd  quark masses as possible.
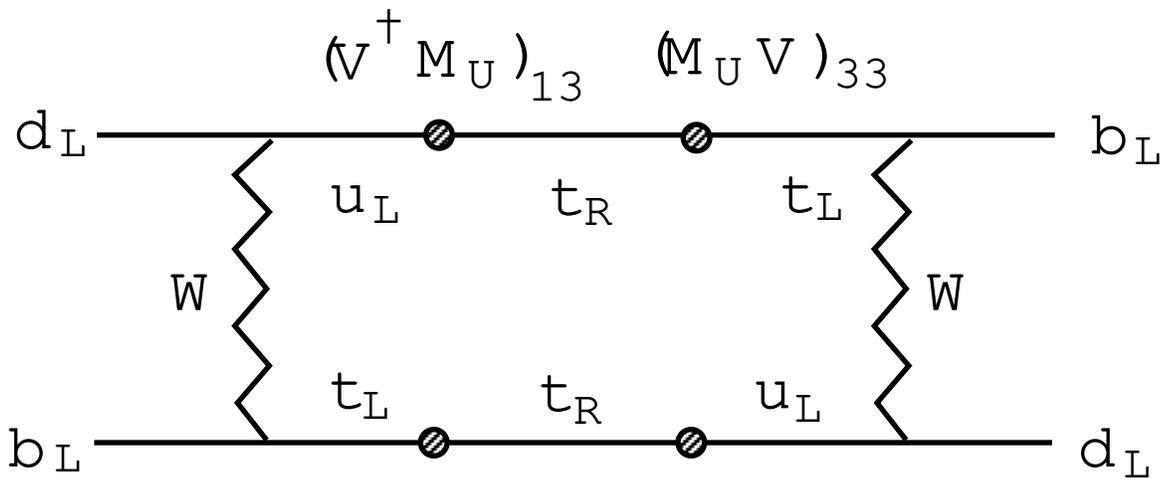
\item[Fig.3] A standard model contribution to $B^0$--$\bar {B^0}$ mixing
calculated in the mass insertion scheme taken here.
\end{description}

\newpage
\begin{thebibliography}{99}
\bibitem{bpr} For reviews, see\\
I.I.~Bigi, V.A.~Khoze, N.G.~Uraltsev,
and A.I.~Sanda, in {\it CP Violation}, ed. C.~Jarlskog (World Scientific,
Singapore, 1989);\\
Y.~Nir, in {\it Proc. of the 20th SLAC Summer Institute
on Particle Physics}, ed. L.~Vassilian, SLAC-REPORT-412 (1992);\\
J.L.~Rosner, in {\it B DECAYS revised 2nd edition},
ed. S.~Stone (World Scientific, Singapore, 1995);\\
Y.~Nir and H.~Quinn, {\it ibid.}
\bibitem{SMB}H.Y.~Cheng, Phys. Rev. {\bf D26}, 143 (1982);\\
A.J.~Buras, W.~Slominsky and H.~Steger, Nucl. Phys. {bf B245},
369 (1984).
\bibitem{KM} M.~Kobayashi and T.~Maskawa, Prog.Theor.Phys. {\bf 49},
652 (1973).
\bibitem{bfac} For a review, see
M.~Artuso, in {\it B DECAYS revised 2nd edition},
ed. S.~Stone (World Scientific, Singapore, 1995).
\bibitem{NFC} G.C.~Branco, Phys. Rev. Lett. {\bf 44}, 504 (1980).
\bibitem{SUSYB} J.-M.~G\'{e}rard, W.~Grimus, A.~Masiero, D.V.~Nanopoulos,
and A.~Raychaudhuri, Nucl. Phys. {\bf B253}, 93 (1985);\\
M.~Dugan, B.~Grinstein, and L.~Hall, Nucl. Phys. {\bf B255}, 413 (1985);\\
G.~Altarelli and P.~Franzini, Z. Phys. {\bf C37}, 241 (1988);\\
T.~Kurimoto, Phys. Rev. {\bf D39}, 3447 (1989).
\bibitem{OKG} Y.~Okada, {\it preprint} KEK-TH-428, KEK Preprint 94-192,
               hep-ph/9502240, February, 1995.
\bibitem{INL} T.~Inami and C.S.~Lim, Prog.Theor.Phys. {\bf 65},
297 (1981).
\end{thebibliography}
\end{document}